\documentclass{emulateapj}

\usepackage{graphics}
\usepackage{rotating}

\newcommand{\kms}{\ifmmode~{\mathrm km~s}^{-1}\else ~km~s$^{-1}~$\fi}
\newcommand{\kmsnospace}{\ifmmode~{\mathrm km~s}^{-1}\else ~km~s$^{-1}$\fi}
\newcommand{\msun}{\ifmmode {M_{\odot}}\else${M_{\odot}}$\fi}
\newcommand\hb{\ifmmode {\mathrm H}\beta \else H$\beta$\fi}
\newcommand{\oii}{\ifmmode{\mathrm [O\,II]} \else [O\,II]\fi}
\newcommand{\oiii}{\ifmmode{\mathrm [O\,III]} \else [O\,III]\fi}
\newcommand{\aox}{\ifmmode{\alpha_{\mathrm ox}} \else $\alpha_{\mathrm ox}$\fi} 
\newcommand{\fcgs}{\ifmmode erg~cm^{-2}~s^{-1[B}\else erg~cm$^{-2}$~s$^{-1}$\fi}
\newcommand{\nh}{\ifmmode{\mathrm N_{H}} \else N$_{H}$\fi}
\newcommand{\nhgal}{\ifmmode{ N_{H}^{Gal}} \else N$_{H}^{Gal}$\fi}
\newcommand{\nhintr}{\ifmmode{ N_{H}^{intr}} \else N$_{H}^{intr}$\fi}
\def\s1254{SDSS\,J1254+0846}

\newcommand{\opteml}{\ifmmode l_{\mathrm 2500\,\AA} \else $~l_{\mathrm 2500\,\AA}$\fi}
\newcommand{\xeml}{\ifmmode l_{\mathrm 2\,keV} \else $~l_{\mathrm 2\,keV}$\fi}
\newcommand{\gax }{{\lower0.8ex\hbox{$\buildrel >\over\sim$}}}
\newcommand{\lax }{{\lower0.8ex\hbox{$\buildrel <\over\sim$}}}

\shorttitle{SDSS~J1254+0846: A Merging Binary Quasar} 
\shortauthors{Green et al.}

\received{Oct 21, 2009}
\accepted{Jan 9, 2010} 

\begin{document}

%% LaTeX will automatically break titles if they run longer than
%% one line. However, you may use \\ to force a line break if
%% you desire.

\title{SDSS~J1254+0846: A Binary Quasar Caught in the Act of Merging}

%% Use \author, \affil, and the \and command to format
%% author and affiliation information.
%% Note that \email has replaced the old \authoremail command
%% from AASTeX v4.0. You can use \email to mark an email address
%% anywhere in the paper, not just in the front matter.
%% As in the title, you can use \\ to force line breaks.

\author{Paul J. Green\altaffilmark{1,2}}
\email{pgreen@cfa.harvard.edu}
\author{Adam~D.~Myers\altaffilmark{2,3},
Wayne~A.~Barkhouse\altaffilmark{2,4},
John~S.~Mulchaey\altaffilmark{5},
Vardha N.~Bennert\altaffilmark{6},
Thomas~J.~Cox\altaffilmark{1,5},
Thomas~L. Aldcroft\altaffilmark{1},
\& Joan M. Wrobel\altaffilmark{6}}

\altaffiltext{1}{Harvard-Smithsonian Center for Astrophysics, 60 Garden
 Street, Cambridge MA 02138}
\altaffiltext{2}{Visiting Astronomer, Kitt Peak National Observatory
  and Cerro Tololo Inter-American Observatory, National Optical
  Astronomy Observatory, which is operated by the Association of
  Universities for Research in Astronomy, Inc. (AURA) under cooperative
  agreement with the National Science Foundation.}
\altaffiltext{3}{Department of Astronomy, University of Illinois at
  Urbana-Champaign, 1002 West Green Street, Urbana, IL 61801-3080}
\altaffiltext{4}{Department of Physics and Astrophysics, University of
  North Dakota, Grand Forks, ND 58202}
\altaffiltext{5}{The Observatories of the Carnegie Institution for Science,
   813 Santa Barbara Street, Pasadena, CA 91101}
\altaffiltext{6}{Department of Physics, University of California,
Santa Barbara, CA 93106}
\altaffiltext{6}{National Radio Astronomy Observatory, P.O. Box O,
  Socorro, NM 87801, USA}

\begin{abstract}
We present the first luminous, spatially resolved binary quasar that
clearly inhabits an ongoing galaxy merger.  SDSS\,J125455.09+084653.9
and SDSS\,J125454.87+084652.1 (SDSS\,J1254+0846 hereafter) are
two luminous $z$=0.44 radio quiet quasars, with a radial velocity difference of
just 215\,km/s, separated on the sky by 21\,kpc in a disturbed host
galaxy merger showing obvious tidal tails.  The pair was targeted as
part of a complete sample of binary quasar candidates with small transverse
separations drawn from SDSS DR6 photometry.  We present follow-up
optical imaging which shows broad, symmetrical
tidal arm features spanning some 75\,kpc at the quasars' redshift.
Previously, the triggering of two quasars during a merger had only 
been hypothesized but our observations provide strong evidence of such an 
event. SDSS\,J1254+0846, as a face-on, pre-coalescence merger hosting two 
luminous quasars separated by a few dozen kpc, provides a unique opportunity 
to probe quasar activity in an {\it ongoing} gas-rich merger. 
Numerical modeling suggests that the system consists of two
massive disk galaxies prograde to their mutual orbit, caught during
the first passage of an active merger. This demonstrates rapid black hole growth during the early stages of a
merger between galaxies with pre-existing bulges.  Neither of the two
luminous nuclei show significant instrinsic absorption by
gas or dust in our optical or X-ray observations, illustrating that
not all merging quasars will be in an obscured, ultraluminous phase.
We find that the Eddington ratio for the fainter component B is rather 
normal, while for the A component $L/L_{\rm Edd}$ is quite ($>$$3\sigma$) high
compared to quasars of similar luminosity and redshift,
possibly evidence for strong merger-triggered accretion.  More such
mergers should be identifiable at higher redshifts using binary
quasars as tracers. 
\end{abstract}

%% Keywords should appear after the \end{abstract} command. The uncommented
%% example has been keyed in ApJ style. See the instructions to authors
%% for the journal to which you are submitting your paper to determine
%% what keyword punctuation is appropriate.


% INTRODUCTION
\section{Introduction}
The origin, growth, and evolution of massive galaxies, and the 
supermassive black holes (SMBHs) that they host, represents a prime
field of study in modern astrophysics. We now know that galaxies
regularly interact and merge \citep{Toomre72}, and that SMBH reside in
the centers of most, if not all galaxies (e.g.,
\citealt{Richstone98}). These two facts alone suggest that binary 
SMBHs should be commonplace. Of course, one or both of the SMBHs in a
binary will only be detectable as quasars when they are actively
accreting. One of the leading proposed mechanisms to trigger strong
accretion (quasar) activity is galaxy mergers (e.g.,
\citealt{Hernquist89,Kauffmann00,Hopkins08}, and references therein),
so merging galaxies with binary quasars should also be common.
\citep{Begelman80} first discussed  binary SMBH
evolution, from galaxy merger to coalescence, as an explanation for
the form and motion of radio jets in Active Galactic Nuclei (AGN).
The `Final Parsec Problem' \citep{Milos03} --
whether the coalescence of a binary SMBH ultimately stalls
 \citep{Milos01}, proceeds to rapid coalescence
(e.g., \citealt{Escala04}), or instead recoils or is ejected (e.g.,
\citealt{Madau04}) -- has important implications for the
detection of gravitational waves and for the spin and demography of SMBHs.  

In a broader context, astronomers hypothesize that ``feedback'' --
whereby dynamical interactions between galaxies trigger accretion onto
their SMBHs -- mediates the tight correlation between galaxy central
black hole masses and the velocity dispersions $\sigma_\star$ of
galaxy bulges ($M_{BH}-\sigma_\star$; \citealt{Ferrarese00,
Gebhardt00}).  The resulting quasars grow in the galaxy
cores until they blow out the very galactic gas that feeds them (e.g.,
\citealt{Granato04}), choking off star formation, and eventually
leading to passive elliptical galaxies \citep{Hopkins07,Kormendy09}.  This
feedback paradigm dovetails with cosmological models of hierarchical
structure formation if quasar activity is induced by massive mergers
(e.g., \citealt{Wyithe02,Wyithe05}).  Major mergers (i.e., those with
mass ratios above $\sim0.3$; \citealt{Shen09}) between gas-rich
galaxies most efficiently channel large quantities of gas inwards,
fostering starbursts and feeding rapid black hole growth.  Deep high
resolution imaging of quasar host galaxies \citep{Bahcall97,
Guyon06,Bennert08} shows strong evidence for fine structure and
tidal tails expected from past gravitational interactions.
Radio-quiet quasar hosts tend to be found in gas-rich galaxy mergers
that form intermediate-mass galaxies, while radio-loud QSOs reside in
massive early-type galaxies, most of which also show signs of recent
mergers or interactions \citep{Wolf08}.  The far-infrared (FIR)
emission of QSOs appears to follow a merger-driven evolution from
FIR-bright to FIR-faint QSOs \citep{Veilleux09a}.

The measured excess of quasars with $\lax$~40\,kpc separations (e.g.,
\citealt{Henn06,Myers07,Myers08}) over the extrapolated large-scale
quasar correlation function may indeed be due to mutual triggering,
but is also debated to arise naturally from their locally overdense
environments \citep{Hopkins08}.  The dynamics and timescales of major
mergers are therefore of the utmost interest.

To date, the merger hypothesis is supported by findings of
spatially-resolved binary active galactic nuclei (AGN) in just a
handful of $z<0.1$ galaxies with one or both of the nuclei heavily
obscured in X-rays (NGC\,6240, \citealt{Komossa03}; Arp\,299,
\citealt{Zezas03}; Mrk\,463, \citealt{Bianchi08}), by the unusual
BL~Lac-type object OJ~287 \citep{Sillanpaa88,Valtonen09}, and perhaps
by X-shaped morphology in radio galaxies (e.g., \citealt{Merritt02,
Liu04,Cheung07}). In addition, COSMOS~J100043.15+020637.2 is known
to contain two AGN resolved at 0.5\arcsec\, ($\sim$1.8\,kpc)
separation in HST/ACS imaging, which have a radial velocity difference
of $\Delta\,v=150$~km/s, and appear to be hosted by a galaxy with a
tidal tail \citep{Comerford09}.

Interest in spatially unresolved systems (spectroscopic binary AGN
candidates) has surged of late, spawned largely from the troves of
SDSS spectroscopy.  The unusual system SDSS~J153636.22+1044127.0
\citep{Boroson09} has a spectrum with two broad line systems
separated by $\Delta\,v=3500$~km/s, and also has a nearby radio
\citep{Wrobel09} and optical \citep{Decarli09}
counterpart. The physical nature of this system has been heavily
debated (e.g., \citealt{Chornock10, Lauer09, Tang09}),
largely because spatially 
unresolved quasars with double-peaked broad emission lines are a quite
common phenomenon (e.g., \citealt{Strateva03}).  Another spatially
unresolved system, SDSS~J092712.65+294344.0, shows 2 broad and 1
narrow emission line system in its spectrum, with
$\Delta\,v=2650$~km/s, sparking discussion about whether it is a
chance superposition \citep{Shields09}, hosts a recoiling SMBH
\citep{Komossa08a}, or is a bound binary SMBH inside a single narrow
line region \citep{Bogdanovic09,Dotti09}.  Quasars with double-peaked
narrow emission lines are relatively common (e.g., there are at least
167 such systems in the SDSS; \citealt{Liu09}). From the DEEP2 Galaxy
Redshift Survey, \citet{Comerford09} found more than a third of Type~2
AGN show \oiii\, line velocities significantly (50--300\kmsnospace)
offset from the redshifts of the host galaxies' stars, arguing that
the most likely explanation is inspiralling SMBHs in merger-remnant
galaxies.  \citet{Smith09} find that about 1\% of (21,592) quasars in
SDSS DR7 have detectable double-peaked \oiii\, emission line profiles.
Only 2 of those appear to be spatially resolved, but the (single-fiber
spectroscopy) sample selection is strongly biased against nuclei with
separations greater than about an arcsecond.

Spatially unresolved systems such as these are relatively easy to find
in large spectroscopic samples.  However, because of the lack of
spatial information, the velocity offsets are open to a variety of
interpretations depending on the relative strength and velocity of
narrow and/or broad emission line systems: small-scale gas kinematics,
asymmetric or thermally inhomogeneous accretion disks, AGN outflows or
jets, recoiling or orbiting SMBHs, or disturbed or rotating narrow
line regions \citep{Smith09}.  Furthermore, spectroscopic samples are
biased against binary AGN that are very close in redshift (i.e.,
unresolved in velocity space), or those that are more widely separated
on the sky.  For instance, in the SDSS spectroscopic survey, the fiber
diameter is 3\arcsec, and the minimum separation of fibers on a plate
is 55\arcsec\, on the sky. So, except in rare cases with multiple
overlapping spectroscopic plates, any binary quasar with separation in
between these two values could only be found from dedicated follow-up
spectroscopy.

Why are spatially-resolved active nuclei in mergers so rare?  First,
they may be heavily shrouded and therefore only detectable as
ultraluminous infrared galaxies. `ULIRGs' have bolometric luminosities
rivalling quasars, and by some ({\em HST } I-band) estimates, as many
as 40\% retain double active nuclei \citep{Cui01}.
A binary fraction in local ULIRGs of at least 40\% is
also consistent with $R$- and $K$-band ground-based data obtained
under $\lax<1\arcsec$ seeing and later confirmed with $H$-band HST
data \citep{Veilleux02,Veilleux06}.
Among dust-reddened quasars, \citet{Urrutia08} found that 85\% show
evidence of merging in images of their host galaxies.  Second,
detectable mergers may be rare simply because the lifetime of the
resolvable-but-unmerged interacting phase is extremely short
\citep{Mortlock99,Foreman09}.  Third, gas-rich major mergers should trace
quasars, and therefore should mainly have occurred 
near the ``quasar epoch'' at higher redshifts  ($z~\gax~1.5$; e.g.,
\citealt{Khochfar01,Wolf03,Silverman05}), where detection of 
extended host galaxy light is challenging.

The prevailing view in the literature (e.g.,
\citealt{Djorgovski91,Kochanek99,Mortlock99,Myers07}) is that the
excess of quasars with small ($<40$\,kpc) separations is evidence for
nuclear triggering in galaxies during dissipative mergers.  According
to Hopkins et al. (2007), the excess measured clustering (e.g.,
\citealt{Henn06,Myers07,Myers08}) indeed represents compelling
evidence for the merger-driven origin of quasars.  However, they also
note that attaching all quasars to moderately rich dark matter
environments in which mergers are most likely to occur is sufficient
to explain the observed excess of binary quasars at $<40$kpc, even if
they are not triggering each other in a bound orbit.  That is, they
just happen to be neighbors where the typical observed velocity
differences could represent $\sim$\,Mpc separations along the line of
sight rather than dynamical velocities, and their properties should be
statistically indistinguishable from those of single quasars.  The
discovery of binary quasars whose hosts are clearly interacting thus
presents rare opportunities to study what merging/triggering really
looks like, and allows for derivation of important quantities
associated with the interaction.

In the next section (\S\ref{sec:discovery}) we discuss the discovery
of SDSS~J1254+0846, a pair of luminous quasars with nearly identical
redshift, hosted by a galaxy merger. Unique among known spatially
resolved systems, SDSS~J1254+0846 is at a scale which suggests it is
an {\it ongoing} merger rather than a relaxed or remnant
system. SDSS~J1254+0846 can thus be used to help study boundary
conditions for gas-rich galaxy mergers.  The description and results
of our deep follow-up multiwavelength imaging and spectroscopy follow
in \S\ref{sec:observations}, including estimates of the black hole
masses and Eddington ratios.  
To verify the plausibility of the merger scenario and determine if
it's possible to infer any properties of the interaction, we  
 compare numerical $N$-body simulations to the
observed properties of the system in \S\ref{sec:numsims}.  In
\S\ref{sec:lens}, we consider the hypothesis that the pair might be
lensed, which we find to be extremely unlikely. We present our
conclusions in \S\ref{sec:conclude}.  Throughout, we assume the
following cosmological parameters for distance-dependent quantities:
$\Omega_{\rm m}=0.3, \Omega_{\Lambda}=0.7$, and $H_0$ = 72
\kms\,Mpc$^{-1}$, which yields an angular size scale of
5.5\,kpc/arcsec at the redshift of our system ($z\sim0.44$).

% Selection and Discovery of SDSS~J1254+0846
\section{Discovery of SDSS~J1254+0846}
\label{sec:discovery}
The objects 
SDSS~J125455.09+084653.9 (SDSS~J1254+0846~A hereafter) and \\
SDSS~J125454.87+084652.1 (SDSS~J1254+0846~B hereafter) 
were targeted as a pair \\ (SDSS~J1254+0846 hereafter) as part of a
complete sample of binary quasar candidates with small transverse
separations drawn from SDSS DR6 photometry (Myers et al. 2010, in preparation). 
A preliminary targeted follow-up campaign of such objects (for DR4) is
discussed in Myers et 
al. (2008). Quasar candidates were selected as having $g < 20.85$ and
either the ``UVX quasar" and/or ``low-redshift quasar" Bayesian
classification flags set in the catalog of Richards et
al. (2009).\footnote{$\textbf{uvxts=1}$ OR $\textbf{lowzts=1}$} These
cuts ensure a high efficiency of quasar pairs in the targets and a
reasonably homogeneous sample over redshifts of $0.4 < z < 2.4$.
Pairs of quasar candidates were then followed up spectroscopically if
they had an angular separation of $3\arcsec$ to $6\arcsec$. To extend
the completeness of the sample as a function of comoving separation,
the sample was also extended to pairs with separations of $6\arcsec$
to $\sim7.7\arcsec$ if neither component had a known redshift at $z >
1.2$.

Following an extensive observational campaign with the R-C
Spectrograph on the Mayall 4-m at Kitt Peak National Observatory and
the Double Spectrograph on the 200-inch Hale telescope at Palomar
Observatory, the sample of target quasar pairs from which SDSS
J1254+0846 was drawn is now complete (again, see Myers et al.~2010, in
preparation). SDSS J1254+0846 itself was discovered to be a binary
quasar on the night of February 11, 2008 and confirmed as such on
February 12, 2008 at Kitt Peak (Myers and Hennawi observing). As was
the case for all candidate binary quasars observed on that run, a
$1.5\arcsec$ by $204\arcsec$ long-slit set at the position angle of
SDSS J1254+0846 was used, allowing both components to be
simultaneously observed. The KPC-10A grating was used yielding a
resolution of $\sim5$\AA~and a wavelength coverage of
$\sim3800$--7800\AA. As the seeing was $\sim1.5\arcsec$ on the nights
in question, and the component separation of SDSS J1254+0846 is
$3.8\arcsec$, care was taken when reducing the data to prevent the
component spectra from merging. The component spectra were reduced
iteratively using xidl's low-redux package (Hennawi, Burles, Schlegel
\& Prochaska; {\tt http://www.ucolick.org/\~\,xavier/LowRedux/}) with
the procedure guided by hand using a boxcar extraction to ensure no
overlap of the spectra.
 
At 1.4~GHz, the host and its quasars are undetected, with flux less
than 2.5\,mJy at a resolution of 45\arcsec\, (250\,kpc) in the NRAO
VLA Sky Survey (NVSS; \citealt{Condon98}) and less than 1\,mJy at a
resolution of 5\arcsec\, (28\,kpc) in the FIRST survey
\citep{White97}).  Following \citet{Ivezic02}, combining the 1\,mJy
upper limit with the $i$-band magnitudes of the quasars, both
components are radio quiet.

As the sample of binary quasar candidates from which it was drawn is
now complete, SDSS J1254+0846 should be unremarkable. On the other
hand, it is the lowest redshift binary quasar currently known and it
has an unusually low $\chi^2$ color similarity statistic (see
Hennawi et al. 2006, Myers et al. 2008) of 0.2, meaning that the
colors of its two components are practically identical. 
Across all 5 bands, their SDSS (PSF) magnitudes differ by
2.28$\pm0.08$ (flux ratio 8.27$\pm$0.61), identical within the errors.
Based on the ($g-i$) vs. redshift (Green et al. 2009) for SDSS quasars,
the optical colors of these quasars are marginally ($\sim0.2$mag) blue
relative to the mean, but consistent with those expected at this
redshift.   The striking feature of SDSS J1254+0846 was discovered 
as we imaged this pair and several others in our Chandra/NOAO joint
program (P. Green, P.I.) to observed binary quasars and their
environments.  Images we obtained on the nights of 18 March 2009
at NOAO's Kitt Peak Observatory with the MOSAIC imager on the 4-m
Mayall telescope (Barkhouse and Myers observing) immediately revealed
bright tidal tails emanating from the quasar pair. The uniqueness of
the system led us to procure further deep imaging and spectroscopy at
other facilities.

% OPTICAL SPECTROSCOPY
\section{Observations}
\label{sec:observations}

\subsection{Optical Spectroscopy}
\label{sec:optspec}

We obtained deeper spectroscopy of both quasars 
on 22 May, 2009 simultaneously through a single 0.9\arcsec\, slit in
seeing of $\sim$0.4\arcsec\, using the IMACS spectrograph on the
Baade-Magellan telescope at the Las Campanas observatory in Chile,.
We centered the slit on QSO B, at a position angle of 61 degrees
to include QSO A.  We used the f/2 camera mode with a 300~lines/mm
grism, giving a wavelength 
range of $\sim$4000 -- 9600\AA\, and dispersion of 1.34\AA/pixel.  We
combined four exposures of 1200s each, and flux-calibrated using the
original SDSS spectrum, except for $>9200$\AA\, where we used the
standard star LTT~3864, rescaled to match the SDSS spectrum, in the
overlap region $8900-9050$\AA. 

\begin{figure}
% \plotfiddle{PSFILE}{VSIZE}{ROTANG}{HSCALE}{VSCALE}{HTRANS}{VTRANS}
%\plotone{flAB.eps}
\plotone{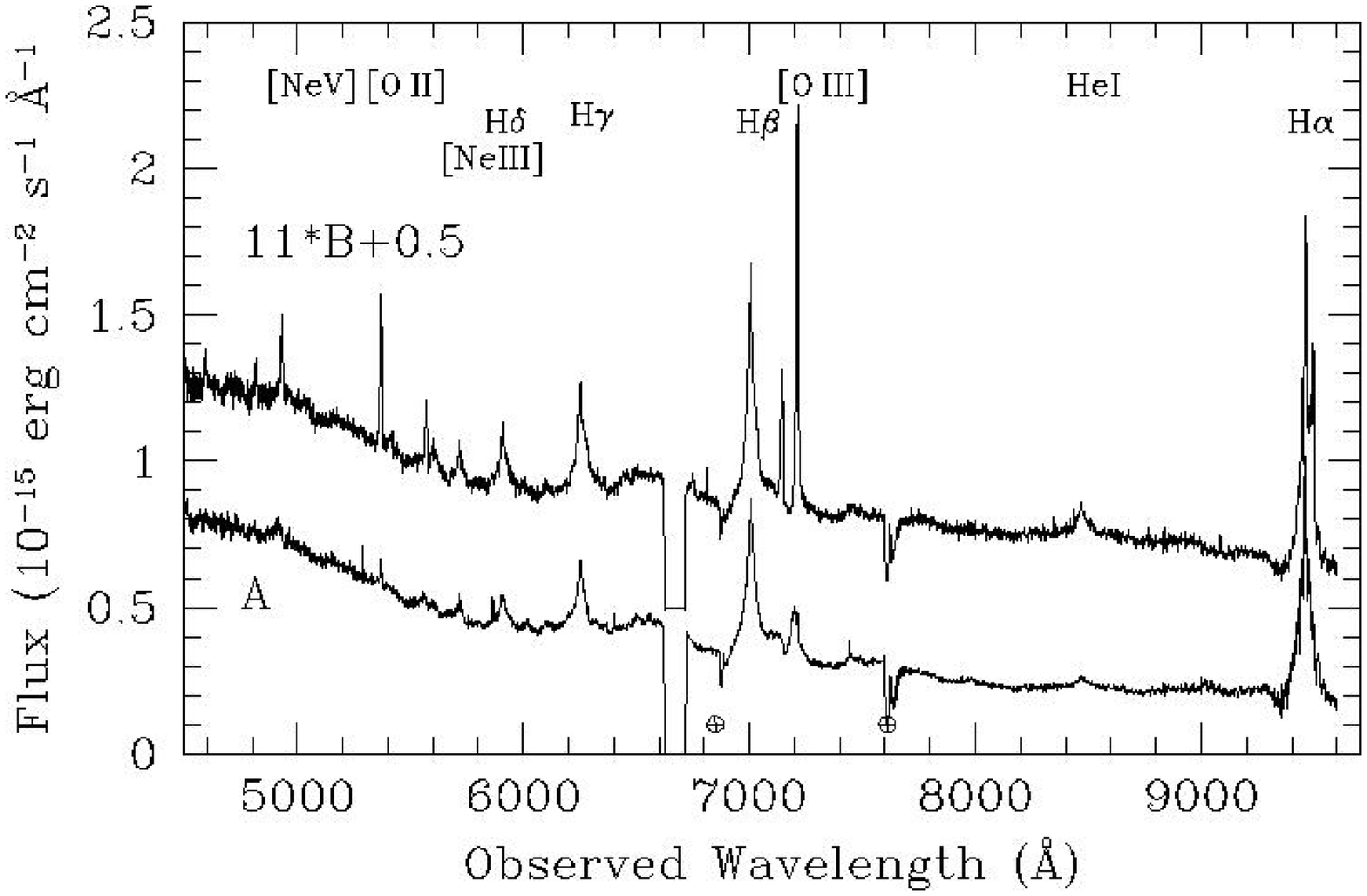}
%\vspace*{-1.25cm}
\caption{Optical spectra. 
The two components' spectra are plotted from 4500 -- 9600\AA, with
the B component scaled up and shifted for clarity.  Despite the factor
of $\sim$11 difference in flux normalization, the redshift, continuum
and broad line shapes are all remarkably similar.  The most striking
difference is in the equivalent widths of the narrow emission lines;
all the forbidden lines of [NeV], [O\,II], [O\,III] are relatively
much stronger in B.  Major emission line species are labeled along the
top.  Major atmostpheric absorption bands are marked with
circumscribed crosses along the bottom.  CCD gaps are evident in both
spectra between $\sim$6600 -- 6700\AA.}
\label{fig:flAB}
\end{figure}

Figure~\ref{fig:flAB} overlays the two spectra, with the B component
scaled up and shifted for clarity.  Despite the factor of $\sim$11
difference in flux normalization, the redshift, continuum and broad
line shapes are all remarkably similar.  The continuum ratio varies
from about 11 near 5000\AA~to about 10 near 8200\AA~(observed frame). 
The discrepancy of 20 -- 30\% between flux ratios in our IMACS
spectroscopy and the SDSS imaging 8.27($\pm$0.61) could be due to
variability of either QSO component between the two epochs.
This is consistent with the somewhat larger discrepancy in the blue,
since QSO variability is known to increase towards shorter wavelengths 
(e.g., \citealt{Wilhite05}).  Some of the difference could also be due
to slit position and alignment.

The most striking difference is in the equivalent widths of the narrow
emission lines.  All the forbidden lines of [NeV], [O\,II], [O\,III]
are {\em relatively} much stronger in B (larger equivalent widths). 
To test whether the spectrum we have extracted for QSO B is 
contaminated by scattered light from A, we  extracted a spectrum on
the other side of QSO A at a distance equal to the separation of A
from B.  No significant spectral features are detectable, so we
conclude that  contamination of the B spectrum by scattered light
from A is negligible.

\begin{figure}
% \plotfiddle{PSFILE}{VSIZE}{ROTANG}{HSCALE}{VSCALE}{HTRANS}{VTRANS}
%\plotone{hbdhbAB.eps}
\plotone{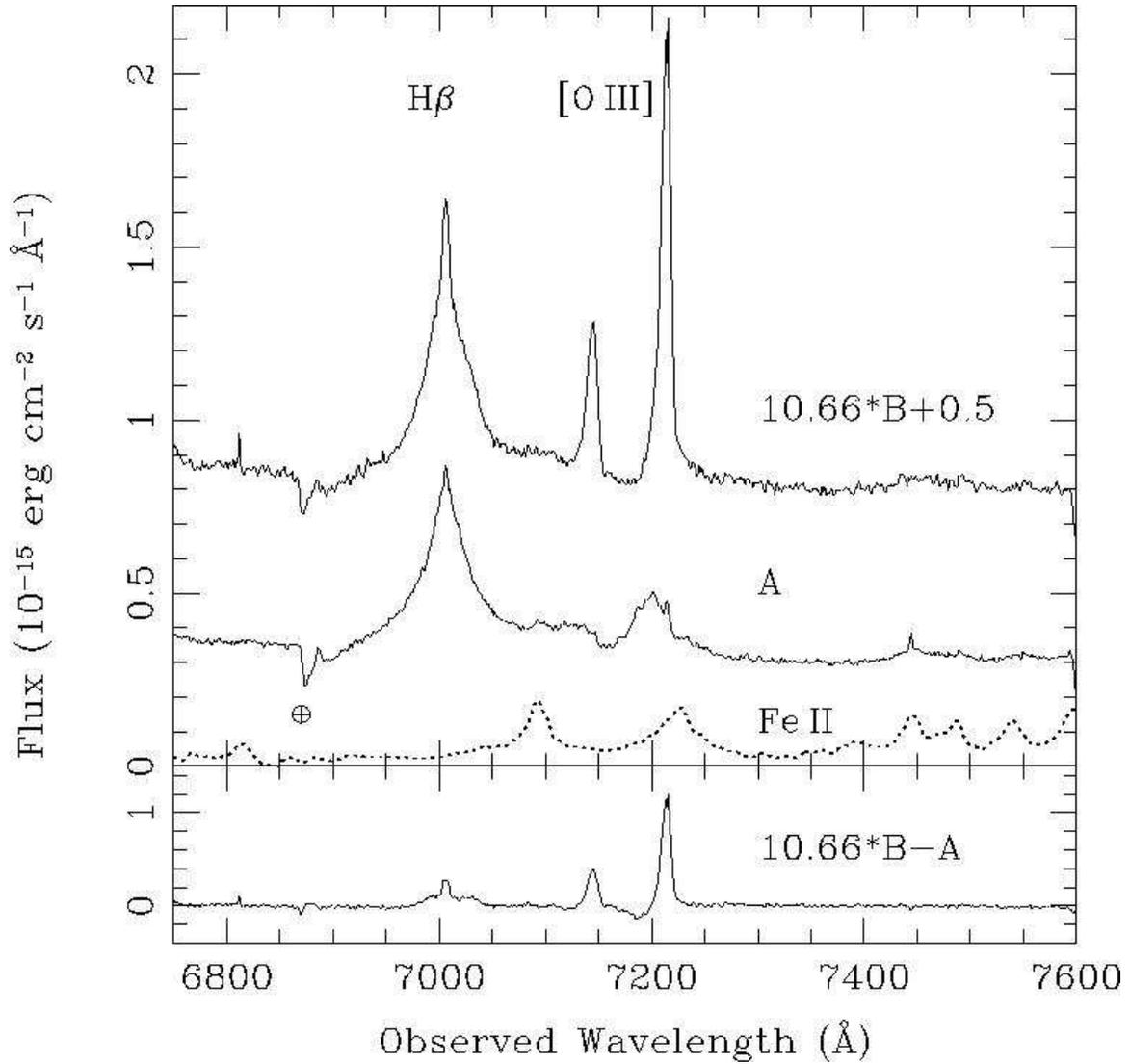}
\caption{Optical spectral features near H$\beta$.
In the upper panel, the two components' spectra are plotted from 6750
-- 7600\AA, with the B component scaled up and shifted for clarity.
The dashed curve is the (arbitrarily scaled) optical Fe\,II emission
template of I\,Zw\,1 (Boroson \& Green 1992).  Such iron multiplet
emission appears to represent at best a minor contribution to either 
spectrum.   The residual of 10.66*B -- A is shown in the bottom panel. 
The flat residuals highlight the similarity of the redshifts,
continuum shape, and broad line profiles.  The apparent trough blueward
of the stronger [O\,III] line is due to the broader, blueshifted
profile of the line in A.}
\label{fig:hbdhbAB}
\end{figure}

\begin{figure}
% \plotfiddle{PSFILE}{VSIZE}{ROTANG}{HSCALE}{VSCALE}{HTRANS}{VTRANS}
%\plotone{hgdhgAB.eps}
\plotone{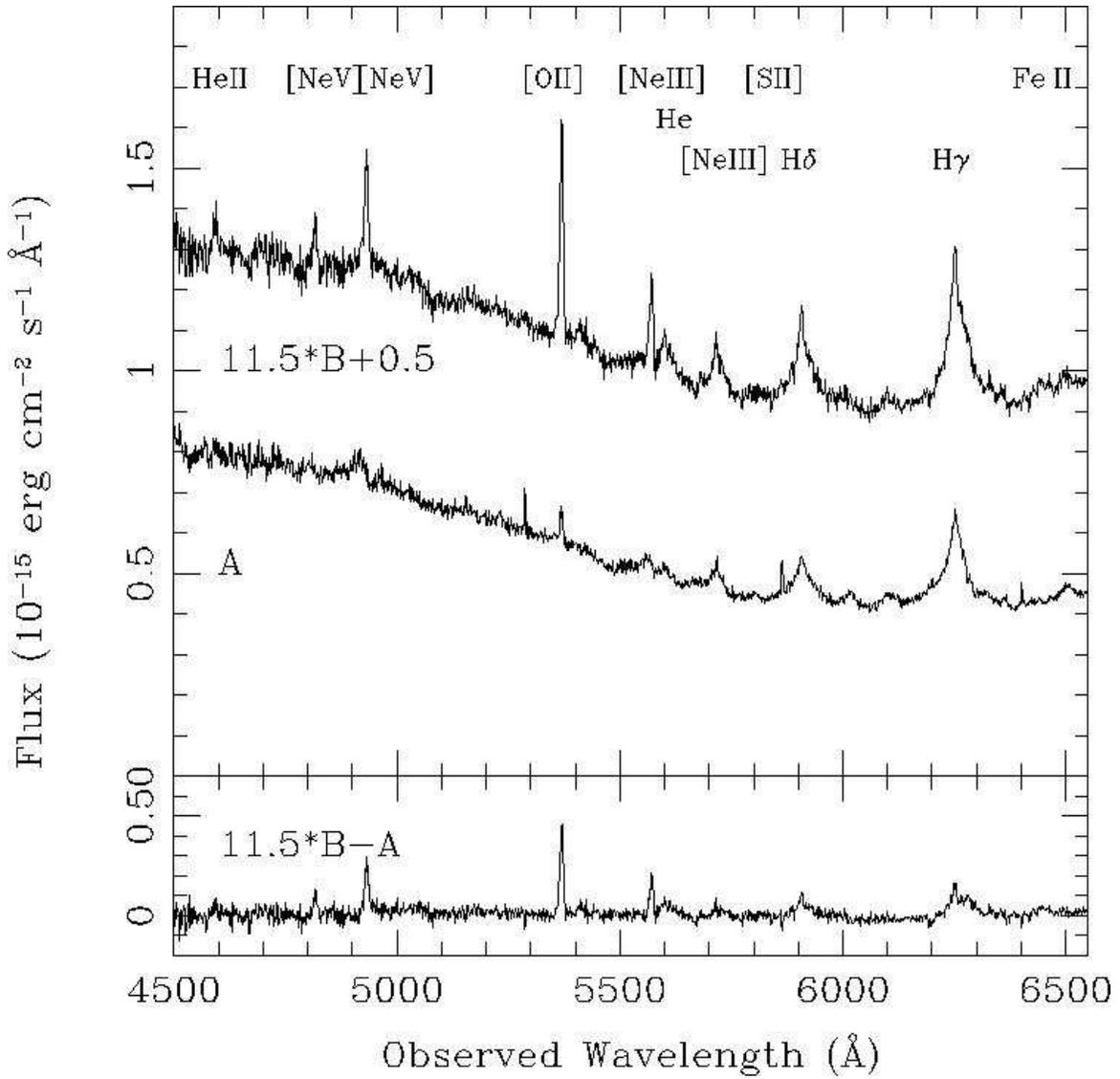}
\caption{Optical spectral features near H$\gamma$.
In the upper panel, the two components' spectra are plotted from 4500
-- 6500\AA, with the B component scaled up and shifted for clarity.
The residual of 11.5*B -- A is shown in the bottom panel.}
\label{fig:hgdhgAB}
\end{figure}

Figures~\ref{fig:hbdhbAB} and \ref{fig:hgdhgAB} highlight the regions
around H$\beta$ and H$\delta$, respectively.  The A spectrum looks
somewhat smoother because of its higher signal-to-noise ratio (S/N).  
The residuals plotted in the lower panels, from simple scaled
subtraction with no velocity shift, are direct evidence that the
redshifts of the two quasars are virtually identical. Separate
cross-correlations of the two spectra against the SDSS median
composite quasar spectrum using IRAF xcsao \citep{Kurtz92}, and
excluding telluric line regions and CCD artifacts yield
$z_A=0.43919\pm0.00028$ ($R$=10.5\footnote{$R$ is the ratio of the
  correlation peak to the amplitude of the asymmetric noise.}) and
$z_B=0.440095\pm0.00011$ ($R$=15.0).  From direct cross-correlation of
the A and B spectra we find a velocity offset for A-B of
-215$\pm$100\,km/s ($R$=6.7), consistent with essentially no 
significant velocity difference.  

The residuals in Figures~\ref{fig:hbdhbAB} and \ref{fig:hgdhgAB}
also illustrate that the most significant differences
between the spectra are in the narrow line components.  However, some
differences are also evident in the broad line regions.  The spectroscopic
differences effectively preclude an interpretation of the pair
as possibly lensed (see \S~\ref{sec:lens} below for further discussion).
Although it is difficult to tell from plots of this rescaled format,
the A/B luminosity ratio is much smaller between the two components in
low-ionization forbidden lines (ratio 1.6$\pm$0.2 in the \oii\,
emission line) than in their continuum emission.  
While the  \oii\, emission cannot be
assumed to be a pure indicator of star formation rate 
in the presence of an AGN \citep{Yan06},
a larger fraction of  \oii\, emission should arise from star
formation in the host galaxy, so the smaller A/B ratio is evidence
that the two nuclei probably share a host. 

% [O\,II] em line flux   B: 4.3e-16   A: 7e-16

We note that if these nuclei were spatially unresolved but with
the same velocity difference and flux ratios, the system would not be
detected in a spectroscopic SDSS search for binary quasars such as that of
Smith et al. (2009).  Direct addition of the two components' spectra
(as if in a single aperture) results in a rather normal-looking quasar
spectrum - the B component is easily subsumed in the A spectrum,
merely highlighting the very peaks of the narrow lines.  On the other
hand, if the tiny velocity difference is due to the serendipitously
small angle of our sightline along the orbital axis, then a different
projection would increase the observed velocity difference, as would
a bound system with smaller physical separation between the nuclei.

\subsection{Optical Imaging}
\label{sec:optim}

\begin{figure}
% \plotfiddle{PSFILE}{VSIZE}{ROTANG}{HSCALE}{VSCALE}{HTRANS}{VTRANS}
%\plotone{Magellan_r1arcmin_CXO_30arcsec_ovlay.eps}
\plotone{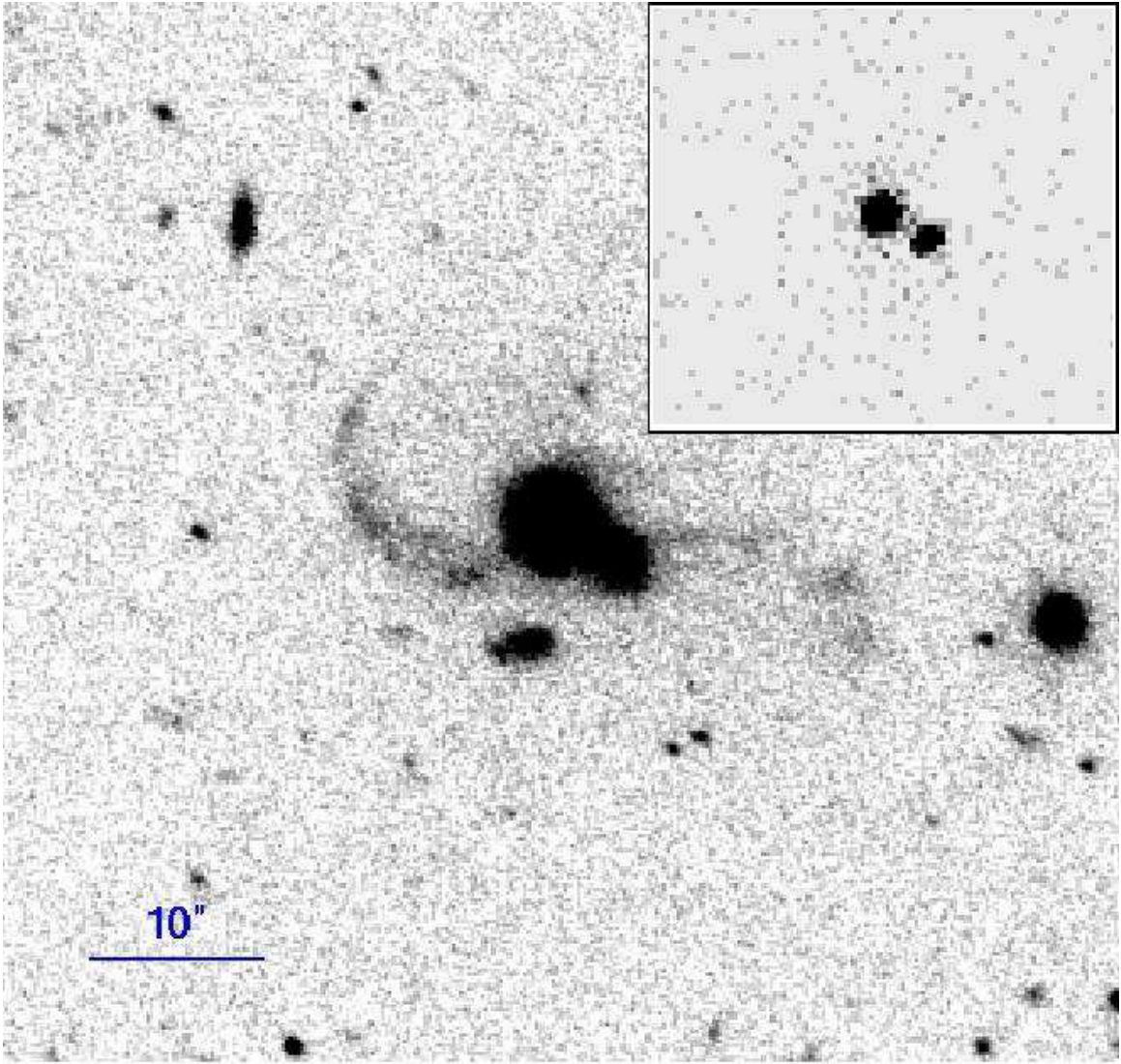}
%\vspace*{-1.25cm}
\caption{Optical and X-ray images of SDSS~J1254+0846.  
This optical image (1\arcmin\, on a side, $N$ up, $E$ to the left)
is the median of four 5min exposures
obtained 2009 May 22 in 1$''$ seeing with an $r$ band filter on the
IMACS camera at the Magellan/Baade telescope at Las Campanas
Observatory in Chile. The two bright quasar nuclei are evident.  The
brighter $A$ component ($r=17.5$) was identified spectroscopically in
the SDSS as a quasar. Discovery spectroscopy of the $B$ component
($r=19.2$) was obtained as part of our binary quasar survey.  This
follow-up IMACS image clearly reveals the tidal arms of a host galaxy
merger.  {\em Inset:} Our Chandra 16\,ksec ACIS-S X-ray image 
(same scale and orientation, but 0.5\arcmin\, on a side) shows
the two nuclei, which both have typical $f_X/f_{opt}$ and power-law
spectral slopes (Green et al. 2009). There is no evidence for
extended emission as might be expected from a host (or lensing-mass)
group or cluster.} 
%\vskip-0.2cm
\label{fig:optximages}
\end{figure}

%OPTICAL IMAGING
On the same night 22 May, 2009 at the Baade-Magellan telescope,
we obtained 20 minutes of imaging (4 exposures of 300sec) in both
Sloan $r$ and $i$ band. The IMACS f/2 camera has 0.2\arcsec\ pixels,
and the seeing was 0.4\arcsec.  We subtracted the CCD bias level and
flattened the field response using averaged dome projector flat images
in each filter as usual.  Images were then co-added with the SWarp
package (Bertin 2006, ver. 2.17.6), and object detection and
measurement were made with SExtractor (Bertin \& Arnouts 1996).
Photometric calibration was performed using dereddened magnitudes from
SDSS DR7 for matching objects in the field.\footnote{We compare SDSS
  {\tt model\_Mag} to SExtractor MAG\_AUTO values.}
The $r$-band image in Figure~\ref{fig:optximages} shows the 2 bright nuclei
of SDSS~J1254+0846 and two symmetrical tidal tails spanning some
75~kpc at the quasar redshift.

\subsection{Chandra X-ray Observations}
\label{sec:cxo}

% CHANDRA IMAGING
We obtained X-ray images of the quasar pair with the Chandra X-ray
Observatory on February 23, 2009 at the ACIS-S aimpoint for 16\,ksec.
The X-ray components are well-resolved by Chandra, and correspond
closely ($<0.2\arcsec$) to their SDSS counterparts.  To avoid
cross-contamination, we extracted the X-ray photons from apertures
corresponding to 90\% of the counts (for 1.5\,keV).  The NE (SW)
components yield 1869 (381) net counts, respectively, in the
0.5-8\,keV range.  We fit an X-ray power-law spectral model 
 $$ N(E) = A\, E^{-−\Gamma}\times \exp [-−N_H^{Gal}\sigma(E) -− N_H^z\sigma(E(1+z_{abs})) ] $$ 
to the counts using the CIAO tool {\it Sherpa}, where $A$ is the
normalization in photons cm$^{-−2}$ sec$^{-−1}$ keV$^{-−1}$ and
$\sigma(E)$ is the absorption cross-section 
\citep{Morrison83,Wilms00}.  We fix $N_H^{Gal}$  at the  
appropriate Galactic neutral absorbing column 
1.9$\times10^{20}$atoms/cm$^{-2}$, and include an intrinsic  
absorber with neutral column $N_H^z$ at the source redshift.  We
group counts to a minimum of 16 per bin and fit using the $\chi^2$
statistic with variance computed from the data.  The best-fit model
for both components is $\Gamma=2.0$ (with 90\% confidence
uncertainties of 0.05 and 0.2 for the NE/SW components, respectively),
and only upper limits to any intrinsic absorption ($\nhintr <$ 2.7 and
7.2 $\times10^{20}$~atoms/cm$^{-2}$, respectively). These values are
quite typical of SDSS quasars \citep{Green09}.  The  2\,keV A/B flux ratio 
is 4.9, somewhat less than the optical flux ratio.  The X-ray to
optical ratio is often parameterized by the X-ray-to-optical spectral
slope  \aox\footnote{$\aox$\, is the slope of a hypothetical power-law from
  2500\,\AA\, to 2~keV; 
  $\aox\,  = 0.3838~{\mathrm log} (\opteml/\xeml)$}, which is
1.41 for A and 1.37 for B.  X-rays in quasars become weaker relative
to optical emission as luminosity increases, and both these quasars
fall along the expected trends \citep{Steffen06,Green09}.  Statistical
tests have shown that the correlation is weaker with redshift, so that
the \aox($L$) relationship is not a secondary effect of quasar
evolution combined with the strong $L-z$ trends of flux-limited quasar
samples. While anecdotal, SDSS~J1254+0846 confirms in a single system
that the observed \aox\ trend with luminosity in quasars is followed
even by quasars at the same epoch and in the same large-scale
environment. Both members of the pair are slightly X-ray bright for
their estimated UV (2500\AA) luminosity.  Our images suggest that the
merger is essentially face-on between massive disk galaxies that are
close to co-planar.  If the quasar accretion disks are reasonably
well-aligned with their galactic host disks\footnote{In radio
  galaxies, there is some general evidence against such alignment
  \citep{Schmitt02}.}, than our sightline may simply avoid the
extinction and redenning associated with a large angle to the line of sight.  

\subsection{Radio Observations}
\label{sec:vla}

% RADIO IMAGING
We observed SDSS\~J1254+0846 with the VLA near transit on UT 2009
September 25 and 27, with net exposure times of 2350~s and 7053~s
respectively, using the DnC configuration under NRAO proposal code 
AG826.  We chose center frequency 8.4601~GHz (8.5~GHz
hereafter) with a bandwidth of 100~MHz for each circular polarization.
Observations were phase-referenced to the calibrator 
J1254+1141 whose positional accuracy was less than 2 mas.  The
switching angle was 3\arcdeg\, and the switching time was 240~s.  
Observations of 3C\,286 were used to set the amplitude scale to an
accuracy of about 3\%.  The data were calibrated using the 2009
December 31 release of the NRAO AIPS software.  Each day's visibility
data for SDSS\,J1254+0846 were concatenated and the AIPS task 
{\tt imagr} was used to form and deconvolve a Stokes $I\/$ image.
Natural weighting was used to optimize sensitivity, giving an angular
resolution at FWHM of 9.4\arcsec\, times 7.2\arcsec\, elongated at
position angle -29\arcdeg.  One source was detected and an
elliptical-Gaussian fit to it, yielding the following integrated flux 
density, position, and 1-dimensional position error: $S = 0.26 \pm
0.03$~mJy, $\alpha(J2000) = 12^{h} 54^{m} 55\fs08$, $\delta(J2000) =
+08\arcdeg 46\arcmin 53\farcs9$, and $\sigma_{\rm VLA} = 0\farcs3$.
The flux density error is the quadratic sum of the 3\% scale error and
the fit residual.  The position error is the quadratic sum of a term
due to the phase-calibrator position error (less than 0.002\arcsec),
the phase-referencing strategies (estimated to be 0.1\arcsec), and the 
signal-to-noise ratio (S/N) (0.3\arcsec).  The source was unresolved
and, given the modest S/N data, the corresponding diameter is less
than the geometric-mean beam width, 8.2\arcsec.

The 8.5 GHz emission from SDSS J1254+0846 has a radio luminosity, $L_R
= \nu L_\nu$, of $1.5 \times 10^{40}$~ergs~s$^{-1}$ and is unresolved,
with a diameter of less than 8.2\arcsec\, (45 kpc).  This scale
encompasses the inner portions of the host galaxy merger, plus quasars
A and B.  Higher-resolution imaging using the Expanded VLA
\citep{Perley09} is needed to localize the emission from each quasar.
In the interim, we note that our Chandra data on quasars A and B
implies a 0.2-20~keV luminosity $L_X$ of $7.9 \times 10^{44}$~ergs~s$^{-1}$.
\citet{Laor08} propose that both active stars and radio quiet AGN owe
their radio emission to similar coronal processes, following
$L_R / L_X \sim 10^{-5}$, where $L_R = \nu L_\nu$ at 5\,GHz.  For a
spectral slope of $-0.5$ \citep{Kellermann94}, the observed 8.5\,GHz
luminosity corresponds to $1.1 \times 10^{40}$~ergs~s$^{-1}$ at
5\,GHz.  Thus $L_R / L_X \sim 1.4 \times 10^{-5}$, just the ratio
expected for the radio-quiet quasars A and B.  This simple, testable
argument suggests that the 8.5 GHz emission arises from both quasars A
and B, without substantial contribution from the extended host galaxy.

\subsection{Black Hole Mass and Eddington Ratio}
\label{sec:mbh}

% BLACK HOLE MASS ESTIMATE
Given our high quality spectra, we can estimate the black hole mass
and Eddington ratios for each quasar.  Most AGN black hole mass
estimators derive from reverberation mapping \citep{Peterson93,Wandel99},
whereby time delays $\tau$ between continuum and broad emission-line
variations are used to deduce the size of the broad line region (BLR).
For single-epoch optical spectra, the continuum luminosity $
L_{\lambda} {\rm (5100\,\AA)}$ can be used as a surrogate for the
broad line region radius $R$ \citep{Koratkar91,Kaspi00,Kaspi05}. 
By combining $\tau$ with the emission-line width (most directly using
\hb), a virial mass for the black hole can be estimated (e.g.,
\citealt{Vestergaard06}). 

We use the {\tt splot} task within IRAF to deblend narrow and broad 
\hb\ above the continuum in each quasar spectrum, and we correct our
\hb\ line width measurements for spectral resolution  and narrow line
contamination following Peterson et al. (2004).  
For quasar A, we measure a H$\beta$ FWHM of 2904$\pm$200\kmsnospace.  The 
log of the continuum luminosity at 5100\AA\, is 45.422 in erg/s.  If
we assume the bolometric correction of 9.26 for 5100\AA\,  
luminosity from \citet{Richards06}, then log\,$L_{\rm Bol}=46.39$.
For quasar B, we measure a H$\beta$ FWHM of 2782$\pm$200\kmsnospace.  The 
continuum luminosity at 5100\AA\, is 44.394 erg/s so log\,$L_{\rm Bol}=45.36$.

%quasar logMbh  logLbol LBol/LEdd logLBol/LEdd
%A   8.461   46.39    0.67      -0.1733
%B   7.796   45.36    0.29      -0.5369

From \citet{McLure04}, we adopt the black hole mass estimator
\begin{equation}
\label{eqn:virial_estimator}
 \log \left({M_{\rm BH,vir} \over M_\odot}\right)
 = 0.672 + 0.61 \log\left({\lambda L_{\lambda} \over 10^{44}\,{\rm 
 erg\,s^{-1}}}\right)+2\log\left({\rm FWHM\over km\,s^{-1}}\right)\,
\end{equation}

% EDDINGTON RATIO
We thereby estimate black hole masses for quasar A and B such that
$ \log \left({M_{\rm BH,vir} \over M_\odot}\right) = 8.46$
and 7.80, respectively.  We estimate uncertainties of about 0.4\,dex
based on \citet{Vestergaard06}.

The Eddington luminosity, assuming a composition of pure hydrogen, is
given by   
\begin{equation}
        L_{\rm Edd}= \frac{4\pi GM_{\rm BH}m_{\rm p}c}{\sigma_{\rm T}} = 1.26\times
 10^{38}\left(\frac{M_{\rm BH}}{M_\odot}\right) {\rm erg\,s^{-1}},
\end{equation}
where $M_{\rm BH}$ is the mass of the black hole, $m_{\rm p}$ is the proton 
mass, and $\sigma_{\rm T}$ is the Thompson scattering cross-section.  
Therefore we find the Eddington ratios $L_{\rm Bol}/L_{\rm Edd}$
for quasars A and B to be 0.67 and 0.29 respectively.
Comparing to \citet{Shen08} for SDSS quasars
in similar ranges of redshift and $L_{\rm Bol}$ (see their Figure~12) ,
the Eddington ratio for quasar B is just 0.6~$\sigma$ above the mean
in log\,$L_{\rm Bol}/L_{\rm Edd}$,
whereas for A the ratio is $\approx 3\sigma$ high.  This could be
evidence that accretion rates are strongly boosted during close
interactions among massive merging galaxies.  

%MERGER SIMULATIONS
\section{Merger Simulations}
\label{sec:numsims}
We can further understand the properties of SDSS\,J1254+0846 via
numerical simulations. 
Galaxy mergers may be a significant triggering mechanism for quasar
activity, and there is growing interest in verifying and understanding
this connection more completely.  Since the majority of theoretical
models (see, e.g., Wyithe \& Loeb 2003, Volonteri et al. 2003, Hopkins et 
al. 2008) associate the most active phase of evolution (and thus most
of the black hole growth) with nuclear coalescence, most quasars are
expected to be hosted by systems where many of the telltale signs of
interaction (disturbed morphology, tidal bridges and tails) no longer
exist, or are difficult to find underneath the glare of the quasar.
In many cases, deep imaging at high spatial resolution
\citep{Dunlop03} but also high signal-to-noise (e.g.,
\citealt{Bennert08}) is required to find evidence for these faint
structures. 

In this context, SDSS~J1254+0846, a pre-coalescence merger with two
observed quasars, provides a unique opportunity to probe the early
phases of the proposed merger/triggering mechanism.  One of the most
powerful insights into this system is via numerical simulation, i.e.,
designing numerical models of the current system which can be evolved,
modified, and compared to the observed system.  We have undertaken
just such a task using numerical techniques that are extensively
detailed in existing literature (e.g., Springel et al. 2005, Cox et
al. 2006, Hopkins et al. 2008, Jonsson et al. 2009).  Briefly, initial
equilibrium disk models are constructed to be representative of disks
at the appropriate redshift.  These models are then initialized on a
prograde orbit and allowed to evolve using the
N-body/SPH code Gadget (Springel 2005) from a distant separation,
through their interaction, to their eventual merger.  The references
provided above include extensive descriptions of these models, and
their generic outcome.

\begin{sidewaysfigure}
% \plotfiddle{PSFILE}{VSIZE}{ROTANG}{HSCALE}{VSCALE}{HTRANS}{VTRANS}
\centering
%\epsscale{0.7}
%\plotone{TJbesfit_14aug09.eps}
\plotone{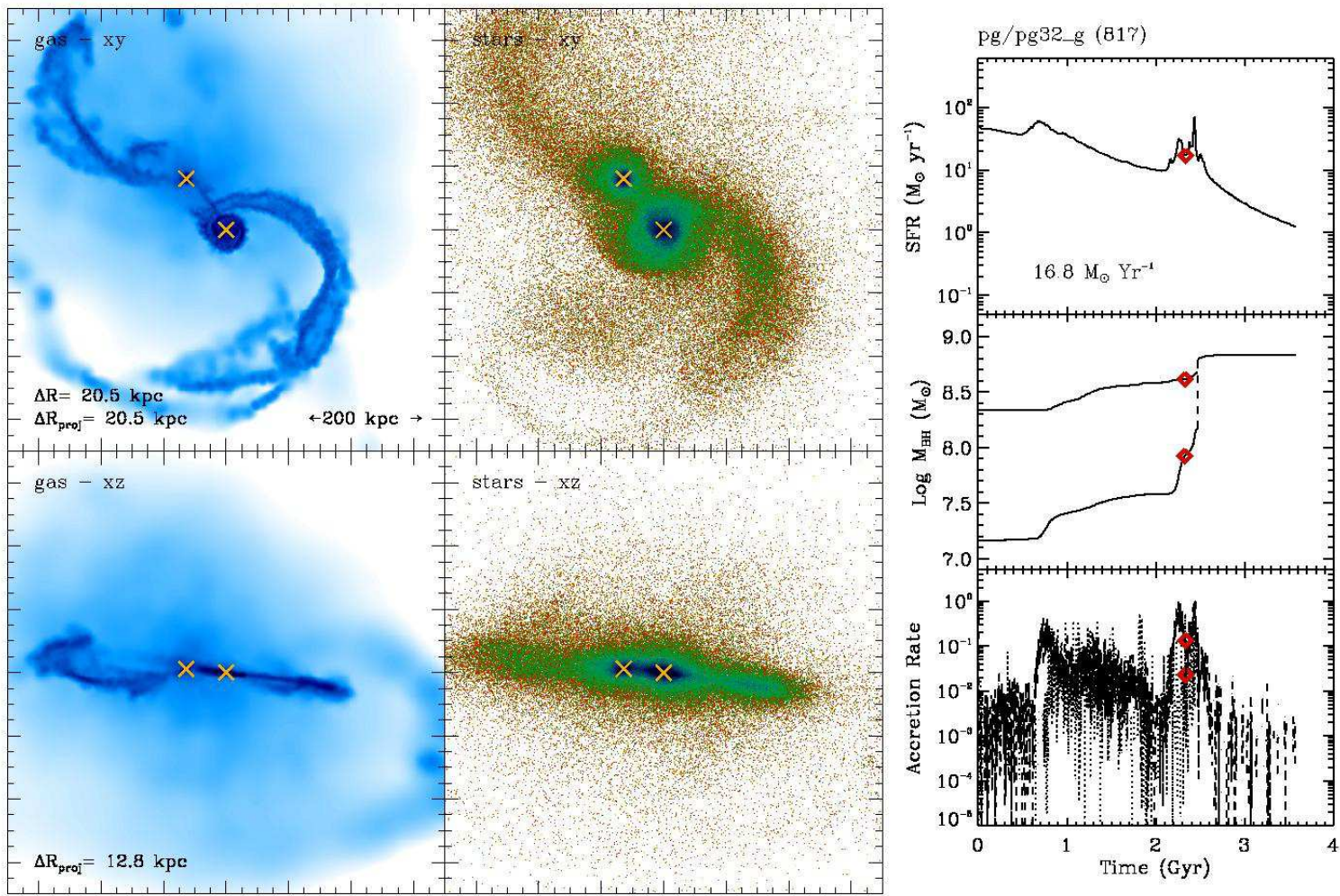}
\caption{Numerical simulation of a merger like SDSS~J1254+0846.
The left panels show the gas distribution, and the middle panels the
stars in the xy and xz planes, for this model of a prograde
merger of two massive disks.  The right panels show the model
star formation rates, nuclear black hole masses and accretion rates
for the two components. The epoch displayed in the images at left,
2.3\,Gyr, is  marked with red diamondsin the plots to the right, 
at which the nuclear separation, the position and extent of the tidal
features (in the xy plane), the black hole masses, and their accretion
rates resemble those of SDSS~J1254+0846.} 
\label{fig:tjsim}
\end{sidewaysfigure}

A representative result of such a numerical modeling experiment is
shown in Figure~\ref{fig:tjsim}, which displays a prograde parabolic
galaxy merger with baryonic mass ratio 2:1, viewed after the second
close passage, but prior to the final coalescence.  This model was
selected owing to both its nuclear separation and the position and
extent of the tidal features showing a remarkable resemblance to SDSS
J1254+0846.  On the other hand, the black hole masses are off by a
factor of $\sim$2 and their accretion rates by a factor of 5-10.

Determining a suitable match proved to be a time-consuming endeavor
which required the analysis of $\sim$200 merger simulations to
isolated the orbits and orientations that best reproduce the observed
tidal features, and the simulation of 8 additional mergers to perfect
this match.  In general, the symmetric tidal features place a
relatively tight constraint on the relative orientation of the disk
spins to that of the orbital plane.  Specifically, the spin-orbit
orientation is required to be less than $\sim$30 degrees.
Furthermore, the relationship between the tidal features and the
nuclear separation demanded a relatively large impact parameter
($R_{peri} \sim 4 R_{disc}$) such that it had a wide first passage,
and a glancing second passage, prior to the final coalescence.
Additional velocity information for both the nuclei and the tails
would place even tighter constraints on these parameters.

The observed tidal features offer a direct means to constrain the
orbital parameters.  However, the set of observed galaxy properties -
specifically the accreting black holes - provide insufficient
information to uniquely determine the properties of the interacting
galaxies.  The matching experiment here only informs us that the
progenitor spiral galaxies are required to be sufficiently large
(i.e., scalelength of the stellar disc $R_disc \gax 4$\,kpc), to produce
the length of the tidal tails, and that they must have contained
pre-existing stellar bulges to match the black hole masses observed at
this early merger stage.  It might be possible to quantify the
properties of the progenitors better with additional information about
the observed system or using additional models, but we caution that
the predicted black hole properties include assumptions about the
initial disk model, including the seed black hole masses, as well as
black hole accretion that occurs well below our model resolution, and
thus additional model-matching is unwarranted until a larger sample of
observed galaxies exists.  In particular, we should not expect
(indeed, we should be skeptical of) a perfect match between the
observed and model-predicted accretion rate onto the supermassive
black holes (and thus also the luminosity of the quasars at any given
time), because stochastic (unpredictable) accretion events appear to
turn ignite nuclear accretion activity at any given time along the
merger sequence in ULIRGs  (e.g., \citealt{Veilleux09b}).
The lower right panel of Figure~\ref{fig:tjsim} makes clear how noisy
the accretion  is expected to be.  

Given the success of our search for a model that matches many of the
properties of the observed system, it is fair to ask what we learn
from such an experiment.  First, we have provided additional evidence
that galaxy mergers are a plausible scenario for triggering quasar
activity.  In fact, nearly all quantities are extracted from the
matched model.  The orbit, the orientation, and the progenitor
galaxies are fully consistent with, and even expected in, a
merger-triggered scenario for quasar formation.  Second, we have
identified a case where the progenitor galaxies participating in the
galaxy merger contain massive bulges and, hence, super-massive black
holes.  While most theoretical models do not currently make testable
predictions about the abundance of such systems during the
hierarchical growth of galaxies and black holes, a larger sample of
such observed systems will certainly motivate additional
investigation.

% LENS INTERPRETATION
\section{A Gravitational Lens?}
\label{sec:lens}
Given the identical redshift, similar colors, and the strong
similarities in continuum slope and broad line profiles between
components A and B, is it possible that the pair is lensed?  It seems
highly unlikey for several reasons.  First, the observed optical
spectra are very different, whereas gravitational lensing should be
essentially achromatic.  Second, the A/B flux ratio is unusually large
for a lensed system.  Third, to achieve the rather wide A/B
separation, a massive lens is expected at intermediate redshift, which
should be sufficiently luminous as to be evident in the images.
Fourth, an intervening lens galaxy should produce some absorption
signature in the quasar spectra.  We discuss these four objections
to a lens interpretation in turn.

Spectral differences between quasar image components are common
even in {\em bona fide} lenses (e.g., Wisotzki et al. 1993; Burud et
al. 2002a,b; Oguri et al. 2005; Sluse et al. 2007).  These differences
are typically explained as the 
effect of either microlensing, or as light path time delays sampling
intrinsic quasar spectral variability.  Even with macrolensing only,
anisotropy in the source may create spectroscopic differences along
the slightly different sightlines (Green 2006, Perna \& Keeton 2009).
While particularly illuminating of source structure, such effects are
expected to be much more subtle than those observed here.  If the pair
is indeed lensed, then microlensing remains the most likely
explanation for the spectral differences.  However,
microlensing-induced spectroscopic differences should primarily affect
emission from the more compact emission regions of the source quasar -
the continuum, or perhaps the broad lines.  The observed spectroscopic
differences are instead predominantly in the narrow lines, whose
emission region is too large (hundreds to thousands of parsecs; e.g.,
\citealt{Motta04,Bennert02}) to be affected by microlensing.

% SORT OUT YOUR ARGUMENTS... MICROLENSING COULD BOOST THE INTRINSIC
% B-LIKE SPECTRUM IN THE CONT/BLR, SO THAT IT LOOKS LIKE A
 
Let us consider the second objection to a lens interpretation,
the unusually large A/B flux ratio in SDSS~J1254+0846.  
While most known {\em bona fide} lens components indeed show smaller
flux ratios, this is quite possibly a selection effect caused by flux
limited surveys from which lens candidates are found.  Furthermore, if the
system were indeed lensed and microlensing were indeed the cause of
the spectroscopic differences described above, the unmicro- (but
still macro-) lensed flux ratio is unlikely to be that observed in the
broadband photometry.  Since microlensing is {\em   least} likely to
affect the narrow line region, the flux ratio of the narrow lines
might more accurately represent the true macrolensed flux ratio. The A/B ratio 
of [O\,III] line flux above the continuum is only about 5.8$\pm$0.4.
For comparison, the mean/median/mode of the A/B $I$-band (HST F814W)
{\em image} flux ratio for 60 lensed quasars in the
CASTLES\footnote{CfA-Arizon Space Telescope Lens Survey information is hosted at
  http://www.cfa.harvard.edu/castles.} database (E. Falco, priv comm)
is 5.4/2.6/1.5, but 8 of the 60 (13\%) have ratios above 9.
Therefore, the large observed image flux ratio in SDSS~J1254+0846 does
not on its own rule out a lens interpretation.

% [OIII] flux from zero (includes continuum!)
%A: 4.0e-14 +/-0.05
%B: 2.2e-15  +/-0.05
% ratio 18.2

% [OIII] flux above continuum
%A: 8.30e-15  +/-0.4
%B: 1.43e-15  +/-0.1
% ratio 

The third objection is that a luminous lensing galaxy is expected
to be visible.  One possibility is that the lensing galaxy of this
system happens to be the tidally disturbed system visible in our
images.    One can ask post-facto, how likely is it to find
an interacting pair of massive spirals with spectacular tidal arms,
well-centered near the mean position of the quasar images?  If the spin of a
galaxy is randomly aligned with the orbit, then only 1/6th of the time
is it aligned within 30 degrees of the orbital spin, and so only
1/36th of the time would both be aligned, such as needed to produce
the observed tails.  It is difficult - perhaps fruitless - to attempt
a further probability calculation for such an alignment, given the
huge parent sample in which this exceedingly rare object was found,
the complex selection effects, and the lack of imaging and spectroscopy of
comparable depths in large statistical samples.  Clearly the most
definitive test of the hypothesis of a tidally disturbed lens galaxy
would be deep  spectroscopy of the faint tails to determine the
redshift of the associated stellar population.

To further investigate the lens hypothesis,
we have run lens models using GRAVLENS 
software (Keeton 2001)\footnote{Software available at 
  http://redfive.rutgers.edu/˜keeton/gravlens/.} 
for a Singular Isothermal Sphere (henceforth SIS) at all intervening redshifts.
For the observed total flux ratio of $\sim$9 we find reasonable
velocity dispersions near 300\,km/s for $0.15<z_{lens}<0.3$. 
From the $I$-band Tully-Fisher relation (Masters et al. 2006), we expect
a $\sigma_V$=300\,km/s galaxy to have absolute $I$-band magnitude
$-20.8\pm0.2$.  For a redshift $z\sim 0.22$, the expected SDSS $i$
band magnitude for such a galaxy is about 19.5, which is about the
same as the fainter quasar.  The expected position of the center of mass
of the lens is determined in the model by the observed flux ratio.
Larger A/B flux ratio means proportionally {\em smaller} distance
from the lens to component A.  In the Appendix, we have attempted to
subtract the A and B nuclear point sources and determine the location,
brightness and significance of any galaxy light between them.  In
summary, while there is some evidence for extended emission around the
quasar nuclei themselves, we find no evidence for significant extended
emission with a centroid consistent with the expected lensing galaxy
position. 

Fourth, a lensing galaxy might be expected to produce detectable
signtaures in the quasar spectrum.  We find insufficient contribution
from stellar emission (either from a $z=0.44$ host or from a putative
intervening lensing galaxy) to create any detectable spectral
features. For an absorber at a plausible
$z_{lens}=0.22$, the more commonly-detected optical/UV
intervening absorption lines (such as Ly$\alpha$ or the Mg\,II doublet
near 2800\AA\, rest) are in the UV.  Detection of the Ca\,II
$\lambda\lambda$3934,3969\AA\, (H\&K) or Na\,I\,D
$\lambda\lambda$5891,5897 absorption would be feasible, but since
these are extremely rare (Wild \& Hewett 2005), a non-detection here
is not useful.  Detailed inspection of the spectra of both A and B
components yields no evidence for any significant absorption lines
that might suggest an intervening (lensing) galaxy.  

X-rays sample intervening gas and dust in all phases.
Our X-ray spectral fit including a $z_{lens}=0.22$ absorption
component, yields an upper limit of 2.2$\times10^{20}$atoms/cm$^{-2}$
from the A component spectrum for an unchanged best-fit continuum
slope $\Gamma=2.01\pm0.08$. The absorption upper limit from B is about
three times weaker.

On balance, we have considered several significant objections to a
lensing scenario, and we believe that despite the nearly identical
redshifts, SDSS~J1254+0846 A and B truly represent a binary quasar.
Our simulations confirm what by all appearances is a merger with
orbits fortuitously close to the plane of the sky, for which the very
similar observed nuclear velocities are much more likely. 

\section{Conclusions}
\label{sec:conclude}

The quasars SDSS~J125455.09+084653.9 (A) and SDSS~J125454.87+084652.1
(B) are within 21\,kpc projected transverse separation at their common
redshift of $z=0.44$, hosted by a galaxy merger showing clear tidal
tail features. The quasar A/B flux ratio is nearly constant across
all 5 SDSS bands, and they show a remarkably small $\sim$2$\sigma$ 
velocity difference of $\sim$200\kms.  We find especially 
strong differences between their narrow emission line equivalent
widths, and their Eddington ratios.  The spectroscopic features of A -
in particular weak \oiii[5007] with evidence for a blueshifted
component - are associated with high accretion rates
\citep{Aoki05,Komossa08b} and outflows.  We suggest that A is very
strongly accreting.  Given its weak, blueshifted \oiii, it would be a
candidate for a broad absorption line quasar, verifiable with UV
spectroscopy of the C\,IV region. A counterargument is that it has a
normal X-ray/optical flux ratio, which is rare in BAL quasars
\citep{Green01,Gallagher06}.

The close coincidence of the positions, colors and redshifts of the
two quasar components raises the suspicion of lensing.  We examine a
variety of counterarguments, most prominently the strong optical
spectroscopic emission line differences, but also the large image flux
ratio, and the absence of either emission or absorption signatures
from an intervening lens galaxy.  Deep spectroscopy of the tidal tails
should prove interesting for a test of the lens model, but also 
for more detailed study of the stellar populations in this
unique system.  Although each counterargument to lensing may have
known caveats, we find the overall evidence to be quite persuasive
that the pair is indeed a binary quasar.  Perhaps the 
strongest argument is simply the association of a binary quasar with a
clear merger of two massive disk galaxies.  While expected under the
merger hypothesis for quasar triggering, we deem the coincidence of such
a system with a lensing configuration to be exceedingly unlikely.

Indeed, a simple explanation for the very similar nuclear radial
velocities, as suggested by the galaxy images and numerical
simulations, is that the merger is apparently along an orbit close to
the plane of the sky.  SDSS~J1254+0846 may represent a rare
system where the orientations of accretion disks in the quasar nuclei
can be constrained by the system configuration as being close to our
line-of-sight (modulo the unknown relative orientations between
accretion disks and galaxy disks).  Such an orientation would be
consistent with the unobscured, Type~1 spectroscopic nature of both
quasars. 

There are strong advantages to studying spatially-resolved binary
quasars (SRBQs) such as these.  SRBQs can be particularly useful when
chosen from well-defined parent samples.  First, such samples probe
{\em ongoing} mergers.  Second, the spatial and velocity information,
especially when combined with well-resolved spectra providing separate
black hole mass estimates, offer more constraints on the properties of
the merging components and the evolution of the merger.  We have found
a good match via numerical merger simulations for the orbit, the
orientation, and the galaxies in this system, showing that it is fully
consistent with a merger-triggered scenario for quasar formation,
where the progenitor galaxies already contain massive bulges.
By selection, SRBQs are likely to be face-on, which makes them ideal for
providing {\em morphological} constraints on merger models via e.g.,  
follow-up with HST or ground-based adaptive optics.  The use of
uniform SRBQ parent samples further allows us to place these systems in
their larger cosmological context, which is crucial if we are to
understand the role of merger-triggered supermassive black hole
accretion, and its relationship to galaxy evolution.

%% Here is the endmatter stuff: Supplementary Info, etc.
%% Use \item's to separate, default label is "Acknowledgements"

\acknowledgments
 Support for this work was provided by the National Aeronautics and
Space Administration  through Chandra\, Award Number GO9-0114A issued
by the Chandra\, X-ray Observatory Center, which is operated by the
Smithsonian Astrophysical Observatory for and on behalf of the
National Aeronautics Space Administration under contract NAS8-03060. 
This paper includes data gathered with the 6.5~meter Magellan
Telescopes located at Las Campanas Observatory, Chile. 
Discovery optical images were obtained at Kitt Peak National
Observatory, National Optical Astronomy Observatory, which is operated
by the Association of Universities for Research in Astronomy (AURA)
under cooperative agreement with the National Science Foundation.  
This research has made use of data obtained from the Chandra Data
Archive and software provided by the Chandra X-ray Center (CXC) in the
application packages CIAO, and Sherpa.  This paper has used data from
the SDSS archive. Funding for the SDSS and SDSS-II has been provided
by the Alfred P. Sloan Foundation, the Participating Institutions,
the NSF, the US Department of Energy, the National Aeronautics and Space
Administration (NASA), the Japanese Monbukagakusho, the Max Planck
Society, and the Higher Education Funding Council for England. The
SDSS website is at http://www.sdss.org/. 

{\it Facilities:} 
\facility{Mayall ()}
\facility{Magellan:Baade ()}
\facility{CXO ()}, 
\facility{VLA} 
\facility{EVLA} 

\dataset [ADS/Sa.CXO#obs/10315] {Chandra ObsId 10315}

\appendix
\section*{Appendix: Surface Photometry}
\label{sec:surfphot}

We used the 2-dimensional galaxy-fitting program GALFIT \citep{Peng02}
to decompose the quasar and host-galaxy light of image components A
and B and to detect and fit underlying extended features.  GALFIT can
simultaneously fit 
one or more objects in an image choosing from a library of functional
forms (e.g., exponential, etc. \citealt{Sersic68, deVauc48}).  For
convolution with the point-spread function (PSF) of the telescope
optics, we first created PSF stars from stars within the image close
to the location of the quasars.  Most stars either had too few counts,
especially in the PSF wings, or were saturated.  We thus created an
artificial Gaussian profile with a FWHM corresponding to that of
non-saturated stars observed in the image.  Qualitatively, the results
are the same as using real star images, with the advantage of zero
noise in the artificial PSF image.  We find that $\gax~95$\% of flux
is removed when we fit stars in the frame, with the residuals due
primarily to  ellipticity in the observed PSF.

For both the $r$ and the $i$-filters we subtracted the sky
background and fitted two PSFs at the locations
of quasar A and B. Simultaneously, we fitted and subtracted closely
neighboring and bright objects.  Due to the saturation of quasar A in our
images, we fixed the flux ratio of the PSFs of quasar A and B to the
ratio given in SDSS-DR7 (using ``psfMag\_r''; ``psfMag\_i'').
We used the SDSS PSF magnitude for quasar B also to calibrate our
results.  Then, we fitted the host galaxies of 
quasar A and quasar B and a potential underlying lensing galaxy with either
a \citet{deVauc48} profile or the more general \citet{Sersic68} profile
\begin{equation}
\Sigma (r) = \Sigma_{\rm eff} \exp \left[- \kappa_n
  \left(\left(\frac{r}{r_{\rm eff}}\right)^{1/n}-1\right)\right]
\hspace{0.2cm} ,
\end{equation}
where $\Sigma_{\rm eff}$ is the pixel surface brightness at the
effective radius $r_{\rm eff}$, and $n$ is the S{\'e}rsic index.
In this generalized form, an exponential disk profile has $n$ = 1, and
a \citet{deVauc48} profile has $n$ = 4.  In general, fitting a
\citet{Sersic68} profile gives more flexibility to the fit, but also adds
an additional free parameter to an already complicated fit, which can
result in an unphysically large S{\'e}rsic index or an
unphysically small effective radius. For this reason, we decided to
fix the S{\'e}rsic index 
to either 4 or 1 (depending on the resulting $\chi^2$) and set the
minimum allowed effective radius to 3 pixels (i.e., the minimum resolvable
size given by the FWHM).  We then chose the best fit based on the
residuals and $\chi^2$ statistics.  Note that we also constructed
masks to exclude tidal structures during the fitting procedure;
however, the effect of the latter on the results is negligible.

In addition to fitting quasar A and quasar B with the PSF model, 
we followed these four approaches:
(1) we fitted one ``joint'' host galaxy at a (starting) location in
between quasar A and B;
(2) we fitted two host galaxies at the locations of quasar A and B;
(3) we fitted two host galaxies at the locations of quasar A and B plus
another galaxy (``lens'') fixed at a position along the A -- B line
expected from a SIS lens model\footnote{More precisely, at the
location along the line between quasar A   and B where the ratio of
separations (quasar B-lens)/(quasar A-lens) equals the A/B flux
ratio based on the SDSS PSF magnitudes.}; 
(4) we fitted two host galaxies at the location of quasar A and B plus
another galaxy (``lens'') close to the location of quasar A.

Only models 2 and 4 yield acceptable fits, with 4 slightly
preferred by comparison of the reduced $\chi^2$ (i.e., taking into
account the larger number of parameters in model 4).
Although it is easier to ``hide'' a lens galaxy near component A
as in model 4, the lens model predicts that the  brighter image ("A")
is outside of the Einstein Radius and is further from the lens. The
fainter image is closer to the lens and is interior to the Einstein
Radius.  Therefore, one result of our experiments is that host galaxies are
required for both A and B.  A single ``joint'' host is unacceptable:
GALFIT instead preferred a host galaxy at the location of either A or
B. The second result is that a third extended component (a putative
``lens'' galaxy in addition to host galaxies for A and B) is preferred
by the fits, but the best-fit position is not the one
predicted by a simple SIS lens model. Fixing an additional 
\citet{deVauc48} profile closer to the fainter quasar results in an
unphysically tiny effective radius.  Freeing the coordinates results
in a second extended component (in addition to the host galaxy) at the
location of quasar B, and one of the two components becomes either
unphysically huge or vanishingly small.  The third (``lens'') galaxy
is allowed only if it is much closer to quasar A.  In
Table~\ref{tab:galfitresults}, we summarize the results
in both $r$ and $i$ filters for the best fit
models, based on residuals and $\chi^2$ statistics.
Based on our experience with fitting quasar host galaxies (and
simulations carried out; \citealt{Bennert10}; see also
\citealt{Kim08}), we conservatively estimate the uncertainties of the
AGN luminosity to 0.2 mag and those of the host galaxies to 0.5 mag.

Note that fitting this system is complicated and our results have to
interpreted with caution.  In general, the decomposition of complex
images in multiple components is a difficult statistical challenge due
the degeneracies involved, and the highly non-linear dependency of the
likelihood on a large number of parameters.  Decomposing quasar
and host-galaxy light is already difficult.  Here, the fitting is further
complicated by the fact that we have two quasars close to each other, one
of which is saturated, with possibly a merging host galaxy and/or
tidal disturbances and/or another underlying galaxy.  Keeping in mind
these cautionary notes, we can conclude the following: There seem to be
two host galaxies at the location of quasar A and B, not just a relaxed
galaxy hosting two quasars.  While there is no evidence for another
galaxy close to the location of quasar B, we cannot exclude the presence
of another galaxy close to the location of quasar A.  However, based
on the flux ratios of QSO A and B, we would expect the lensing galaxy
to fall closer to QSO B. Thus, our modelling rules out significant
extended emission with a centroid that is consistent with the expected
lensing galaxy position.  Deep images of higher signal-to-noise and
smaller PSF are warranted, either from  HST or using adaptive optics
from the ground.

%% Put the bibliography here, most people will use BiBTeX in
%% which case the environment below should be replaced with
%% the \bibliography{} command.

\clearpage

\begin{deluxetable}{ccccccc}
\tabletypesize{\scriptsize}
\tablecolumns{6}
\tablewidth{0pc}
\tablecaption{Results from GALFIT fitting}
\tablehead{
\colhead{Filter} & \colhead{PSF quasar A} &
\colhead{PSF quasar B} & \colhead{Host quasar A} & \colhead{Host quasar B} &
\colhead{``lens''}\\
& mag & mag & mag & mag & mag\\
\colhead{(1)} & \colhead{(2)} & \colhead{(3)}  & \colhead{(4)} &
\colhead{(5)} & \colhead{(6)}}
\startdata
r & 17.15 & 19.50 & 17.52 & 19.33 & \nodata \\
  & 17.15 & 19.50 & 17.98 & 19.33 & 18.23\\
i & 17.07 & 19.36 & 18.44 & 19.85 & \nodata \\
  & 17.07 & 19.36 & 18.75 & 19.49 & 18.37\\
\enddata
\tablecomments{
GALFIT host galaxy fit results.
+Col. (1): Filter: SDSS r or SDSS i.
Col. (2): SDSS PSF magnitude of quasar A.
Col. (3): SDSS PSF magnitude of quasar B.
Col. (4): Best-fit magnitude of quasar A host galaxy.
Col. (5): Best-fit magnitude of quasar B host galaxy.
Col. (6): Best-fit magnitude of third ``lens'' galaxy close to A.
}
\label{tab:galfitresults}
\end{deluxetable}

\end{document}